\documentclass[12pt]{article}

\usepackage{graphicx}

\newcommand{\1}{1\!\!\!\bot}

\begin{document}
\begin{center}
{\Large\bf \boldmath  Infrared Propagators in MAG \\[2mm]
and Feynman gauge on the lattice} 

\vspace*{6mm}
{Tereza Mendes$^{a,b}$, Attilio Cucchieri$^a$, Axel Maas$^{c}$ and 
Antonio Mihara$^d$ }\\      
{\small \it 
            $^a$ Instituto de F\'\i sica de S\~ao Carlos, 
Universidade de S\~ao Paulo, \\
Caixa Postal 369, 13560-970 S\~ao Carlos, Brazil \\
            $^b$ DESY, Platanenalle 6, 15738 Zeuthen, Germany \\
            $^c$ Institute of Physics, Karl-Franzens University Graz, 
                 Universit\"atsplatz 5, A-8010 Graz, Austria \\
            $^d$ Instituto de Ci\^encias Exatas e Tecnologia,
                 Universidade Federal do Amazonas, \\
                 R. Nossa Senhora do Ros\'ario 3863,
                 69100-000 Itacoatiara, Brazil 
}
\end{center}

\vspace*{6mm}

\begin{abstract}
We propose to investigate infrared properties of gluon and ghost
propagators related to the so-called Gribov-Zwanziger 
confinement scenario, originally formulated for Landau and Coulomb 
gauges, for other gauges as well. We present results of our investigation
of $SU(2)$ lattice gauge theory in the maximally Abelian gauge (MAG), 
focusing on the behavior of propagators in 
the off-diagonal (i.e.\ non-Abelian) sector. 
We also comment on our preliminary results for general linear
covariant gauges, in particular for Feynman gauge.
\end{abstract}

\vspace*{6mm}

\section{Introduction}

Important features of quark and gluon confinement in QCD are believed 
to be closely related to the behavior of gluon and ghost propagators 
in the infrared limit. One must notice, however, that the study of 
infrared properties of these propagators must be performed by 
nonperturbative methods and at a fixed gauge.
The Gribov-Zwanziger confinement scenario 
\cite{Gribov:1977wm,Zwanziger:1991ac}
--- proposed for Landau and 
Coulomb gauges --- provides predictions for gluon and ghost
propagators in the infrared limit, which may be tested by lattice
simulations and by nonperturbative analytic methods such as 
Dyson-Schwinger equations.
In particular, a suppressed infrared gluon propagator $D(p^2)$ is predicted,
with $D(0) = 0$. The latter statement implies maximal violation of reflection 
positivity for the gluons, a result that may be viewed as an indication of
gluon confinement. (Note that it suffices to have violation of reflection
positivity, not necessarily maximal.)
At the same time, the infinite-volume limit favors gauge configurations on 
the boundary region known as the first Gribov horizon, where the smallest 
nonzero eigenvalue $\lambda_{min}$ of the Faddeev-Popov matrix ${\cal M}$ 
goes to zero. As a consequence, the ghost propagator $G(p^2)$ --- which is 
obtained from ${\cal M}^{-1}$ --- should be infrared-enhanced, introducing 
long-range effects in the theory. These, in turn, would be responsible for the 
color-confinement mechanism.

The nonperturbative study of infrared propagators may be
carried out from first principles in lattice simulations, taking into
account that the true infrared behavior is however obtained only at 
large enough lattice volumes.
Considerable effort has been dedicated to investigations of the above
predictions for the Landau gauge, considering very large lattice sizes
(see e.g.\ \cite{Cucchieri:2007md}). The status of these studies is 
discussed in \cite{Cucchieri}. 
Here we propose to test similar predictions for the propagators
as applied to the lattice implementation of other gauges, to try to 
gain a unified understanding of the mechanism of confinement and its 
manifestations. We consider the maximally Abelian gauge (MAG) and the 
linear covariant gauges, in particular Feynman gauge. 

In the case of the linear covariant gauges, which include and generalize
Landau gauge, some studies suggest that the Gribov-Zwanziger confinement 
mechanism may apply to the complete class of such gauges
\cite{Alkofer:2003jr,Sobreiro:2005vn}.
(A recent study of Dyson-Schwinger equations for Feynman gauge has been 
presented in \cite{Aguilar:2007nf}.)
On the other hand, for MAG, the usual confinement scenario is
based on the concepts of Abelian dominance and of dual
superconductivity \cite{Kronfeld:1987ri}. Nevertheless, one might argue 
that a modified Gribov-Zwanziger scenario would likely hold in MAG for 
the non-Abelian directions in gauge-configuration space.
A study of the Yang-Mills Lagrangian restricted to the (MAG) Gribov region 
by addition of a horizon function with Gribov parameter $\gamma$
has recently been carried out for $SU(2)$ gauge theory in \cite{Capri:2008ak}.
As pointed out in that reference and also by other groups, the infrared 
behavior of propagators in MAG may be modified by the presence of ghost 
and gluon condensates of mass dimension two. An example of such objects 
is the ghost condensate $v$ \cite{Schaden:1999ew,Capri:2005tj},
related to the breakdown of a global $SL(2,R)$ symmetry. 
This quantity is expected to modify the symmetric and anti-symmetric
components of the (off-diagonal) ghost propagator.
In particular, a nonzero value for $v$ corresponds to nonzero 
anti-symmetric components of the ghost propagator.
In Section \ref{mag} we present results of our lattice studies of pure 
$SU(2)$ theory in MAG. 
(The implementation of gauge fixing for MAG on the lattice is 
straightforward.)
We consider gluon and ghost propagators, the ghost condensate $v$ mentioned
above and the smallest eigenvalue of the Fadeev-Popov matrix.
Our preliminary results have also been presented in \cite{Mendes:2006kc}
and \cite{Mihari:2007zz}.
We note that the bounds recently introduced for studying gluon and
ghost propagators on large lattices in Landau gauge 
\cite{Cucchieri:2007rg,Cucchieri:2008fc} may be written
also for other gauges.

Contrary to the case of MAG, the technical aspect of fixing the
linear covariant gauges on the lattice is still not a settled issue.
We comment on our recent proposals for gauge-fixing methods for these
gauges in Section \ref{feynman}.

\section{Infrared propagators in MAG}
\label{mag}

On the lattice, for the $SU(2)$ case, the MAG is obtained (see e.g.\ 
\cite{Bornyakov:2003ee}) by minimizing the functional
\begin{equation}
S \;=\; - \frac{1}{2 d V} \sum_{x,\mu} Tr
\left[ \sigma_3 U_{\mu}(x) \sigma_3 U^{\dagger}_{\mu}(x) \right]\,.
\end{equation}
At any local minimum one has that the Faddeev-Popov matrix, defined as
\begin{eqnarray}
\!\!\!\!\!\!\!\! \sum_{b y} M^{ab}(x,y) \gamma^b(y) \!\!\!\! &=& \!\!\!\! \sum_{\mu}
\gamma^a(x) [ V_{\mu}(x) + V_{\mu}(x-e_{\mu}) ]
          \, + \,2 \, \{ \gamma^a(x-e_{\mu}) [ 1 - 2 (U_{\mu}^0(x))^2] 
               \nonumber \\[3mm]
\!\!\! & & \!\!\! - \, 2\, \sum_{b} \gamma^b(x-e_{\mu}) [ \epsilon_{ab} U_{\mu}^0(x) U_{\mu}^3(x) +
           \sum_{cd} \epsilon_{ad} \epsilon_{bc} U_{\mu}^d(x) U_{\mu}^c(x) ] \}
\, ,
\end{eqnarray}
is positive-definite. Here the color indices take values
$1, 2$ and we follow the notation
$U_{\mu}(x) = U_{\mu}^0(x) \1 + i\, {\sigma}^a {U}_{\mu}^a(x)$ and
$V_{\mu}(x) = (U_{\mu}^0(x))^2+(U_{\mu}^3(x))^2-(U_{\mu}^1(x))^2-(U_{\mu}^2(x))^2$,
where $\sigma^a$ are the 3 Pauli matrices.
Notice that (as in Landau gauge \cite{Cucchieri:2005yr}) this matrix
is symmetric under the simultaneous exchange of color and space-time
indices.
Using the relation $U_{\mu}(x) = \exp{[- i a g_0 A_{\mu}(x)]}$
one finds (in the formal continuum limit $a \to 0$) the standard
continuum results \cite{Bruckmann:2000xd} 
for the stationary conditions above and for $M^{ab}(x,y)$.

We have considered four values of $\beta$ (2.2, 2.3, 2.4, 2.512)
and lattice volumes up to $40^4$. Data for a larger lattice, of
volume $56^4$, have also been recently produced, but are not fully
analyzed yet. We include data for the ghost propagator at this volume
in Fig.\ \ref{cond} below, for comparison.
 
Our results for the gluon propagators are in agreement with the 
study by Bornyakov et al.\ \cite{Bornyakov:2003ee}:
we see a clear suppression of the off-diagonal propagators compared to
the diagonal (transverse) one, supporting Abelian dominance.
We have fitted our data for the various gluon
propagators (at all values of $V$ up to $40^4$ and for $\beta = 2.2$),
obtaining the following behaviors. For $D(p^2)$ (transverse) diagonal,
our data favor a Stingl-Gribov form 
\begin{equation}
D(p^2) \;=\; \frac{1+d\,p^2}{a+b\,p^2+c\,p^4}\,,
\label{SG}
\end{equation}
with a mass $\,m = \sqrt{a/b} \,\approx \, 0.72\,GeV\,$.
Note that the above equation corresponds to a pair
of complex-conjugate poles $z$ and $z^*$. We can thus write
$z = x + i y$ with $x = b/(2 c) \approx 0.32 \, GeV^2$ and
$y = \sqrt{a/c - x^2} \approx 0.47 \, GeV^2$. Let us recall that
in the case of a Gribov-like propagator these two poles are
purely imaginary.
For $D(p^2)$ transverse off-diagonal our best fit is
of Yukawa type, i.e.\ $D(p^2) = 1/(a+b\,p^2)\,$, with
a mass $\,m = \sqrt{a/b} \,\approx \, 0.97\,GeV\,$.
Finally, the longitudinal off-diagonal gluon propagator
is best fitted by
$D(p^2) = 1/(a+b\,p^2+c\,p^4)$ (i.e.\ also of Yukawa type) 
with a mass $\,m = \sqrt{a/b} \,\approx \, 1.25\,GeV\,$.
As expected from Abelian dominance, the mass is larger
in the off-diagonal case.

\begin{figure}
\includegraphics[width=0.45\textwidth]{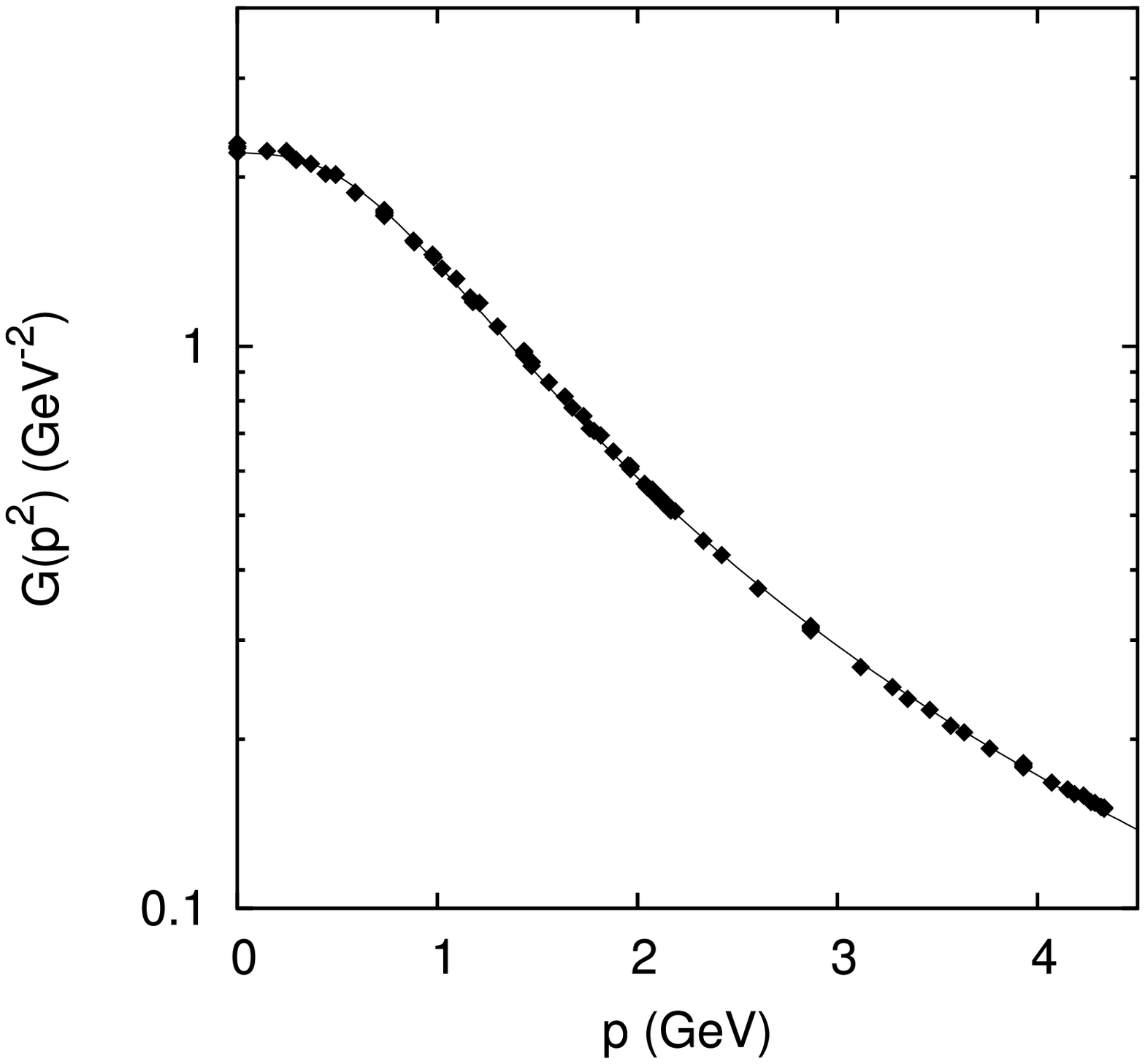} \hspace{5mm}
\includegraphics[width=0.48\textwidth]{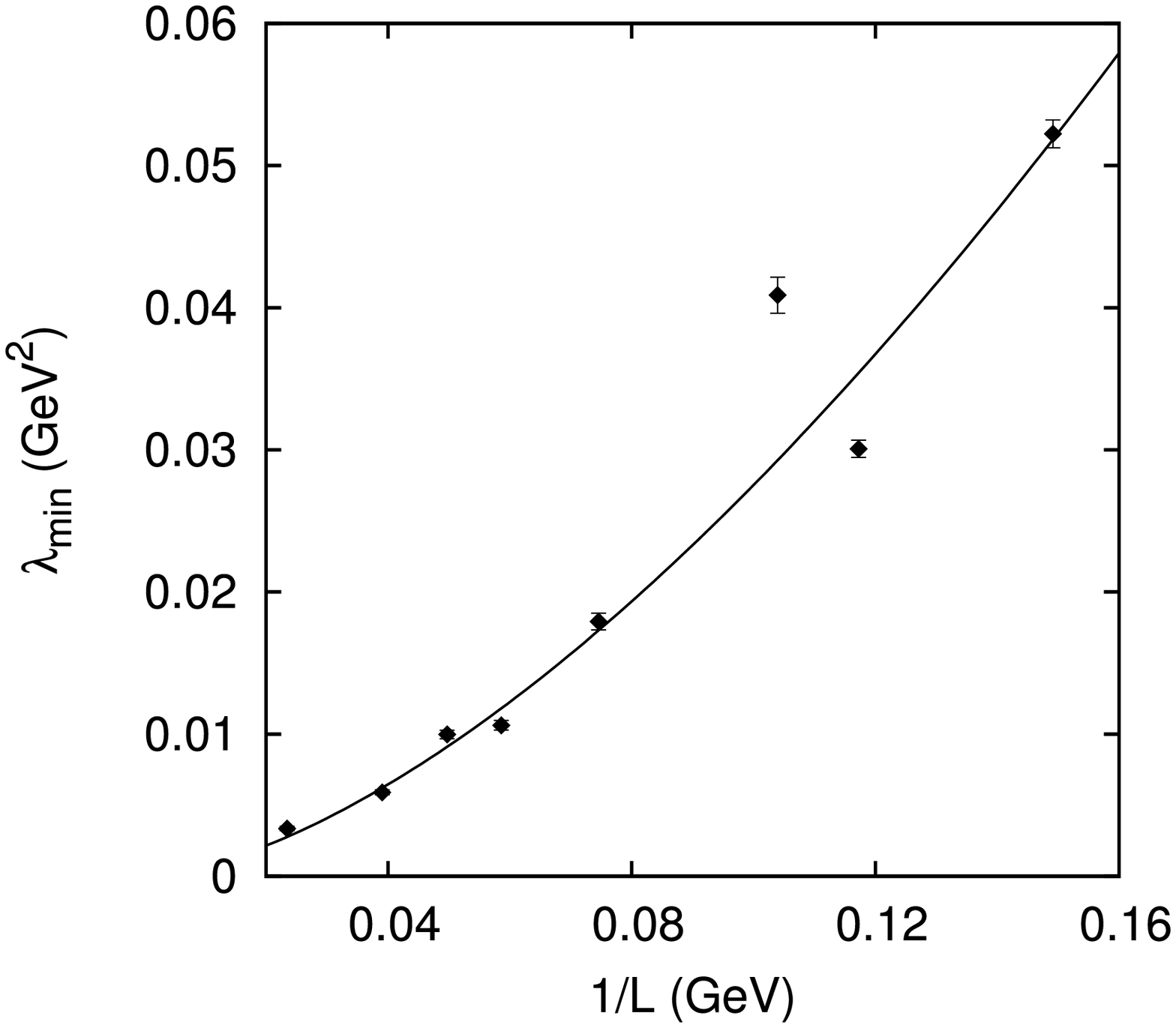}
\caption{Left: plot of $G(p^2)$ as a function of improved $p$
for lattice volumes $V = 16^4$, $24^4$, $40^4$ and $\beta = 2.2$.
Right: plot of the smallest eigenvalue of the Faddeev-Popov operator, 
as a function of the inverse linear size of the system.
}
\label{ghost}
\end{figure}

In Fig.\ \ref{ghost} (left) we show our data for the 
ghost propagator $G(p^2)$, as a function of an improved momentum $p$
(see Ref.\ \cite{Ma:1999kn}).
The data show little volume dependence at small $p$.
(Note that, contrary to Landau gauge, here we can evaluate the ghost 
propagator at zero momentum.)
We see no sign of an enhanced IR propagator.
We have fitted our data (at $\beta = 2.2$),
obtaining a behavior of the type (\ref{SG}) above
with
$a = 0.45(1)\,GeV^2\,,$
$b = 1.1(3)\,,$
$c = 0.73(30)\,GeV^{-2}\,,$
$d = 2.1(9)\,GeV^{-2}\,.$
Thus, we see a Stingl-Gribov fit with mass
$\,m \,\approx \, 0.6\,GeV\,$ and complex poles given by
$x\approx 0.75$, $y\approx 0.22$.

We next consider (see Fig.\ \ref{ghost}, right)
the smallest eigenvalue $\lambda_{min}$ of the Faddeev-Popov matrix.
We have looked at $\lambda_{min}$ for several lattice volumes and values 
of $\beta$ as a function of $1/L$. The data are fitted
to $\,a\,(1/L)^b\,$ with $\,b = 1.6(1)$, showing that $\lambda_{min}$
vanishes more slowly than $(1/L)^2$ (Laplacian).
This may explain why we do not see a diverging ghost propagator at zero
momentum even at rather large lattice volumes \cite{Cucchieri:2008fc}.

Following the analysis done in Landau gauge \cite{Cucchieri:2005yr}, 
we consider the anti-symmetric off-diagonal ghost propagator
$ \langle \, | \, \epsilon_{ab} G^{ab}(p^2) /2 \, | \,\rangle $
rescaled by $L^2 / \cos\left(\pi \,\widetilde{p}_{\mu}\, a/L\right)$,
as a function of the (unimproved) momentum $p$ for all lattice volumes 
and $\beta$ values considered.
The data show nice scaling for all cases considered.
The data at $\,V = 40^4$ and $\,\beta = 2.2\,$ can be fitted by
$\,\Phi(p) = (a+b\,p/L^2)(p^4+v^2)\,$ with
$a = 0.0026(7)\,GeV^2\,,$
$b = 32.6(7)\,GeV^{-1}\,$
and
$v^2 = 1.7(1)\,GeV^{4}\,.$
We thus have a rather large ghost condensate
$\,v \approx 1.3 \,GeV^2\,$, but we cannot be sure that it survives
in the infinite-volume limit, since the overall constant $a$ might be null.
We can also fit data at several $V$'s and $\,\beta$'s for
$\,\Phi(p^2)\,$ as a function of $p$ and $L$ (see Fig.\ \ref{cond}, right).
We obtain $\Phi(p) = (a+b\,p/L^2)(p^4+v^2)\,$ with
$a = 0.0033(6)\,GeV^2\,,$
$b = 35.8(5)\,GeV^{-1}\,$
and
$v^2 = 1.87(8)\,GeV^{4}\,.$
We note that the fit parameters change little with the (physical)
lattice volume.
In fact, data obtained recently for a larger lattice volume,
$56^4$, are seen to fall nicely on top of the fit done for the
smaller volumes, as seen in Fig.\ \ref{cond} (right).
\begin{figure}
\includegraphics[width=0.45\textwidth]{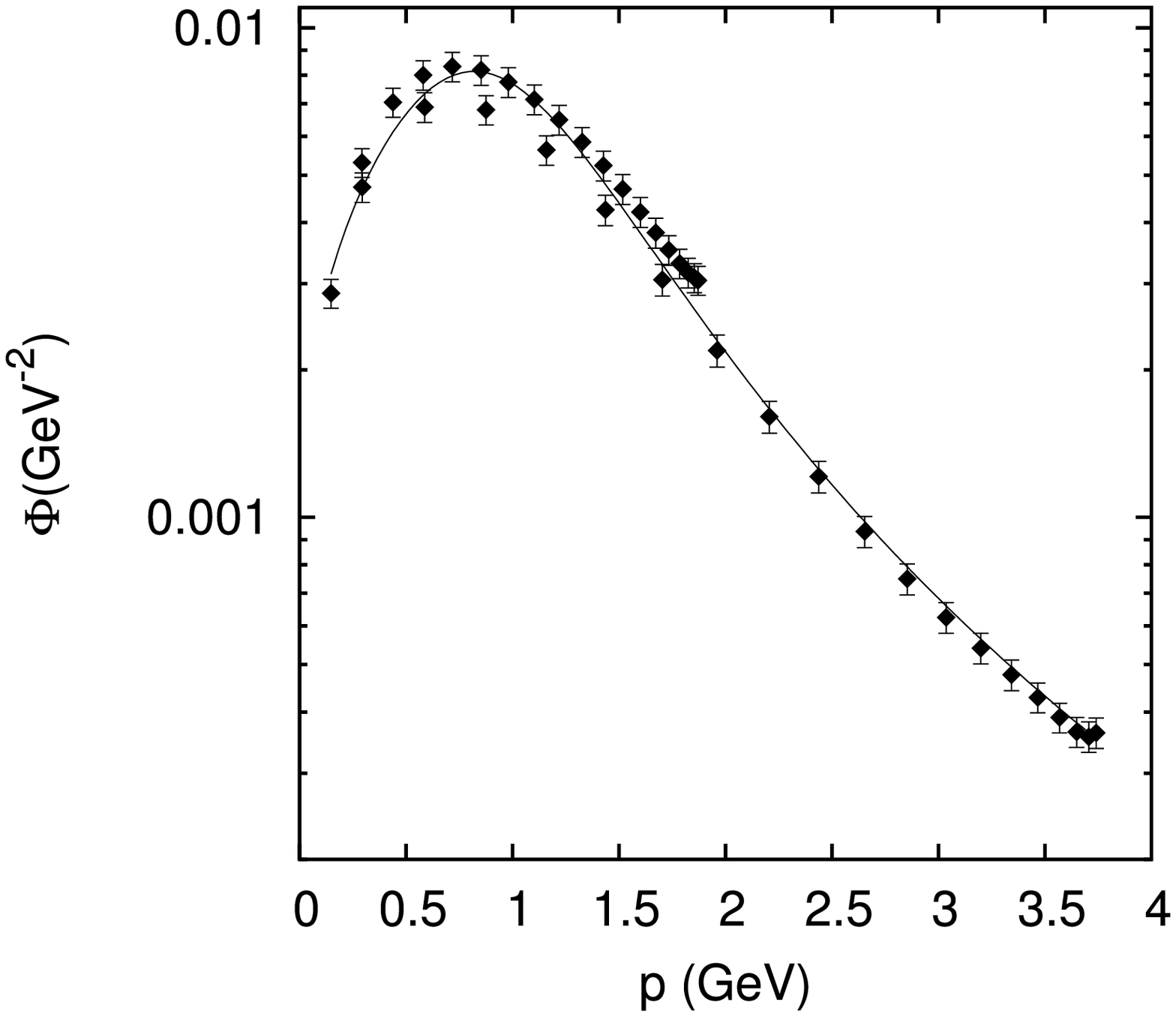} \hspace{5mm}
\includegraphics[width=0.45\textwidth]{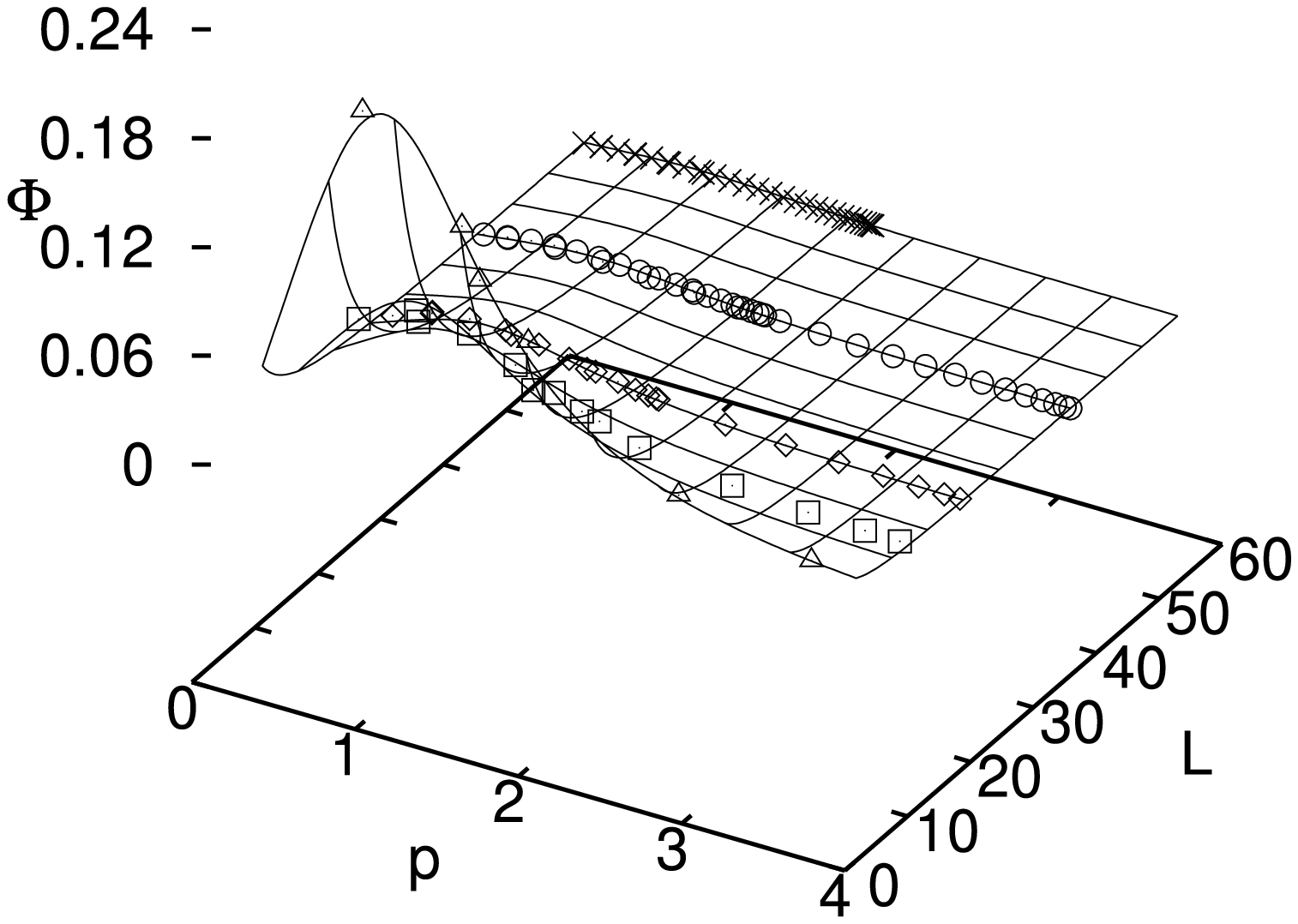}
\caption{Left: plot of the quantity $\Phi(p^2) =
L^2 / \cos\left(\pi \,\widetilde{p}_{\mu}\, a/L\right)
\langle \, | \, \epsilon_{ab} G^{ab}(p^2) /2 \, | \,\rangle $
as a function of $p$ for lattice
volumes $V = 8^4$, $16^4$, $24^4$, $40^4$ and $\beta = 2.2$.
Right: plot of $\Phi(p^2)$ as a function of $p$ and $L$.
Data for a larger volume, $56^4$, are also included here.}
\label{cond}
\end{figure}

We have also investigated possible effects of Gribov copies on our 
results, by considering the difference between our standard gauge fixing
(using the stochastic overrelaxation algorithm \cite{Cucchieri:1995pn}) 
and the so-called smearing method \cite{Hetrick:1997yy}. The effects are 
found to be of the order of the statistical error.

\section{Linear covariant gauges}
\label{feynman}

\begin{figure}[t]
\centerline{\includegraphics[width=0.5\textwidth]{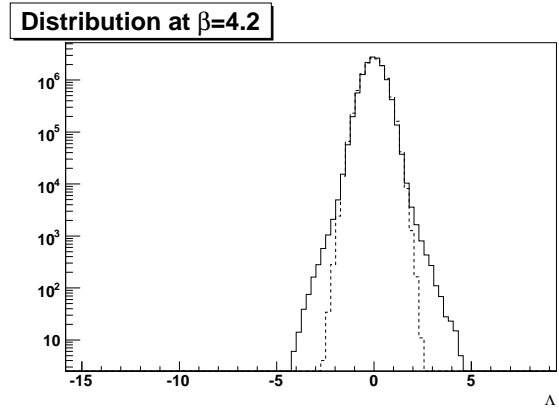}}
\caption{Distribution of $\partial {A'}^a_{\mu}(x)$
(solid line) compared with a Gaussian of width $\sqrt{\xi}$
(dashed line).}
\label{dist}
\end{figure}

As mentioned in the Introduction, gauge-fixing to linear covariant
gauges (other than Landau gauge) on the lattice is a challenge.
More precisely, the gauge condition is given by 
\begin{equation}
\partial_{\mu} A^a_{\mu}(x)\;=\; \Lambda^a (x)
\label{landau}
\end{equation}
for real-valued functions $\Lambda^a (x)$. As opposed to the
case of Landau gauge --- for which $\Lambda^a (x)=0$ and the gauge is
fixed by minimizing a simple functional of the gauge-transformed
links --- in the general case no such functional exists \cite{Giusti:1996kf}.
The solution to this problem presented in \cite{Giusti:1996kf}, based on the
consideration of a modified gauge-fixing condition for the minimizing functional,
may be affected by
spurious minima and it leads to an altered form of the Faddeev-Popov matrix.
We propose to consider
a class of gauges on the lattice that coincides with the perturbative
definition of linear covariant gauges in the formal continuum limit.
Our method is based on a three-step process.
Instead of minimizing a functional of $\Lambda^a (x)$ directly, we first 
fix the gauge to Landau gauge, i.e.\ the transformed gauge fields satisfy 
$\partial_{\mu} A^a_{\mu}(x)=0$. Then we determine a transformation 
$\phi^b(x)$ such that
\begin{equation}
{A'}^a_{\mu}(x)\;\equiv\; A^a_{\mu}(x)\,+\,D^{ab}_{\mu}\phi^b(x)
\end{equation}
satisfies Eq.\ (\ref{landau}).
Finally, we repeat the procedure for several functions $\Lambda^a (x)$ with a
Gaussian distribution of width $\sqrt{\xi}$. The case $\xi=1$ corresponds
to Feynman gauge. The resulting distribution of $\partial {A'}^a_{\mu}(x)$ 
is shown for $\xi=1$ in Fig.\ \ref{dist}, in comparison with the original
(Gaussian) distribution taken for $\Lambda^a (x)$.
We see that the expexted distribution is fairly well reproduced.

Our preliminary results were presented in \cite{Cucchieri:2008zx}.
We are currently investigating an alternative method for fixing these gauges.

\vskip 3mm
\section{Acknowledgments}
We thank S. Sorella for helpful discussions.
A.\ C.\ and T.\ M.\ were partially supported by FAPESP 
and CNPq. The work of T. M. was also supported by the
A. von Humboldt Foundation. 
A. Maas was supported by the DFG under grants number MA 3935/1-1 
and number MA 3935/1-2 and by the FWF under grant number P20330.
A.\ Mihara was partially supported by CNPq.

\end{document}